\documentclass[acmtog,screen,nonacm]{acmart}

\acmSubmissionID{1388}

\usepackage{booktabs}
\usepackage{multirow}
\usepackage{graphicx}
\usepackage{amsmath}
\usepackage{stfloats}
\usepackage{enumitem}
\usepackage[capitalise]{cleveref}
\usepackage{colortbl}

\usepackage{xcolor}

\graphicspath{{./images/}}

\newcommand{\tablestyle}[2]{\setlength{\tabcolsep}{#1}
                            \renewcommand{\arraystretch}{#2}
                            \centering
                            \footnotesize}

\definecolor{graycolor}{gray}{.9}

\copyrightyear{2025}
\acmYear{2025}
\setcopyright{acmlicensed}
\acmConference[SA Conference Papers '25]{SIGGRAPH Asia 2025 Conference Papers}{December 15--18, 2025}{Hong Kong, China}
\acmBooktitle{SIGGRAPH Asia 2025 Conference Papers (SA Conference Papers '25), December 15--18, 2025, Hong Kong, China}
\acmDOI{10.1145/3757377.3763869}
\acmISBN{979-8-4007-2137-3/2025/12}

\begin{document}


\title{Neural Visibility of Point Sets}

\author{Jun-Hao Wang}
\affiliation{
  \department{Wangxuan Institute of Computer Technology}
  \institution{Peking University}
  \country{China}
}
\email{wangjh24@stu.pku.edu.cn}
\authornote{Joint first authors.}

\author{Yi-Yang Tian}
\affiliation{
  \department{Wangxuan Institute of Computer Technology}
  \institution{Peking University}
  \country{China}
}
\authornotemark[1]

\author{Baoquan Chen}
\affiliation{
  \department{School of Intelligence Science and Technology}
  \institution{Peking University}
  \country{China}
}

\author{Peng-Shuai Wang}
\affiliation{
  \department{Wangxuan Institute of Computer Technology}
  \institution{Peking University}
  \country{China}
}
\email{wangps@hotmail.com}
\authornote{Corresponding author.}

\begin{abstract}
  Point clouds are widely used representations of 3D data, but determining the visibility of points from a given viewpoint remains a challenging problem due to their sparse nature and lack of explicit connectivity.
  Traditional methods, such as Hidden Point Removal (HPR), face limitations in computational efficiency, robustness to noise, and handling concave regions or low-density point clouds.
  In this paper, we propose a novel approach to visibility determination in point clouds by formulating it as a binary classification task.
  The core of our network consists of a 3D U-Net that extracts view-independent point-wise features and a shared multi-layer perceptron (MLP) that predicts point visibility using the extracted features and view direction as inputs.
  The network is trained end-to-end with ground-truth visibility labels generated from rendered 3D models.
  Our method significantly outperforms HPR in both accuracy and computational efficiency, achieving up to \emph{126 times} speedup on large point clouds.
  Additionally, our network demonstrates robustness to noise and varying point cloud densities and generalizes well to unseen shapes.
  We validate the effectiveness of our approach through extensive experiments on the ShapeNet, ABC Dataset and real-world datasets, showing substantial improvements in visibility accuracy.
  We also demonstrate the versatility of our method in various applications, including point cloud visualization, surface reconstruction, normal estimation, shadow rendering, and viewpoint optimization.
  Our code and models are available at \url{https://github.com/octree-nn/neural-visibility}.
\end{abstract}

\begin{teaserfigure}
  \centering
  \includegraphics[width=\linewidth]{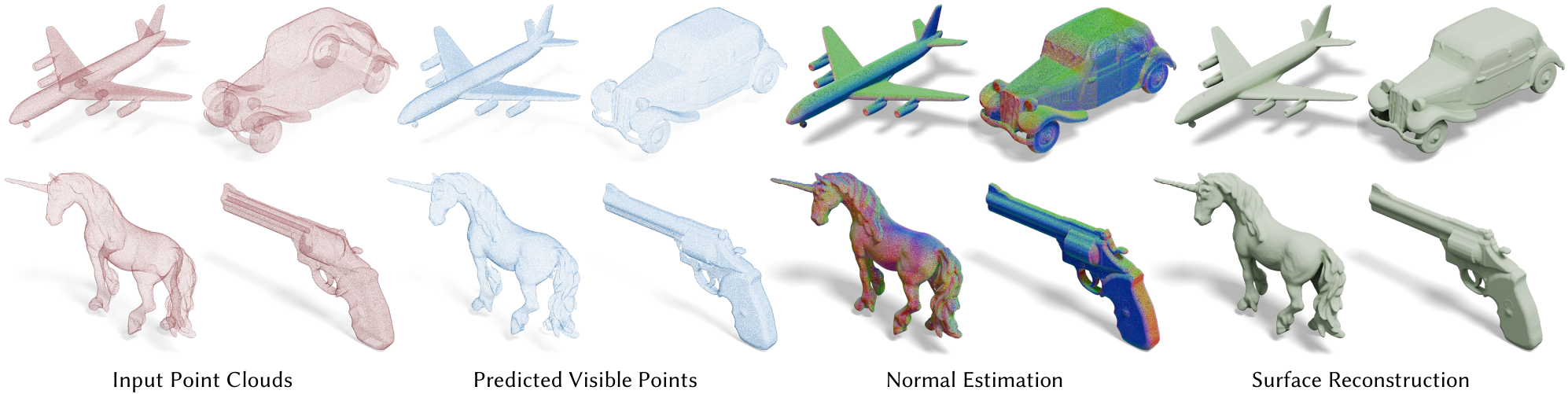}
  \caption{Our method learns to directly predict the visibility of each point in a point cloud from a given viewpoint. It accurately determines visibility in real time, even in the presence of noise and varying point cloud densities, thereby enabling a wide range of applications.
  From left to right, the visualizations show the input point clouds, the corresponding predicted visible points from a given viewpoint, the estimated normals based on the predicted visibility, and the final reconstructed meshes using Poisson surface reconstruction.
}
  \label{fig:teaser}
\end{teaserfigure}




\maketitle
\section{Introduction}\label{sec:intro}

Point clouds are among the most widely used representations of 3D data in computer graphics, computer vision, and robotics.
They comprise a set of discrete points in 3D space, possibly augmented with additional attributes such as normals or colors, to capture the geometry and appearance of 3D objects or scenes.
However, directly rendering or visualizing point clouds poses challenges, as determining the visibility of points from a given viewpoint is not straightforward and can often be ambiguous.

Mathematically, a point is infinitesimal in size, and the probability of one point being precisely occluded by another from a given viewpoint is nearly zero.
A common approach to resolving visibility in point clouds involves reconstructing surfaces from input point clouds~\cite{Kazhdan2006,Hoppe1992} and performing visibility tests on these reconstructed surfaces.
However, surface reconstruction itself remains an open problem~\cite{Berger2017} and is often computationally expensive, particularly for large, noisy, or unoriented point clouds.
An alternative line of methods, known as Hidden Points Removal (HPR)~\cite{Katz2007,Katz2015,Mehra2010,katz2025hpro}, directly determines point visibility without requiring surface reconstruction by transforming the point cloud relative to a sphere centered at the viewpoint and then computing its convex hull.
While HPR offers a more direct approach, it is also computationally demanding for large point clouds due to the convex hull computation.
Additionally, it performs poorly in the presence of noise, concave regions, or low-density point clouds, limiting its applicability in practical scenarios.
There are also methods that project point clouds into pixel space and use neural networks to predict the visibility of the projected points~\cite{Pittaluga2019,Song2021}.
The accuracy of these methods is often limited by image resolution and point cloud density.

In this paper, we formulate the visibility determination problem in point clouds as a binary classification task and train a neural network to predict the visibility of each point from a given viewpoint.
Our key insight is that the visibility of a point cloud is an inherently ill-posed problem without knowledge of the underlying surface.
However, we can leverage the power of deep learning to learn the necessary prior information from data, overcoming the limitations of conventional hand-crafted methods like HPR.
Given a point cloud, we first extract point-wise features using a 3D U-Net~\cite{Ronneberger2015} built upon octree-based CNNs~\cite{Wang2017}.
Subsequently, we employ a lightweight, shared MLP to predict the visibility of each point, taking the extracted point-wise features and viewpoint information as inputs.
During training, the whole network is supervised using ground-truth visibility labels generated by rendering the corresponding 3D models from which the point clouds are sampled.
Once trained, the network can efficiently extract features for an unseen point cloud in a single forward pass.
The visibility of each point is then determined by evaluating the shared MLP, incorporating the viewpoint information.

Our method offers several advantages over existing approaches.
First, it is highly efficient and can predict the visibility of point clouds in real-time.
The network is readily parallelizable on GPUs, and the extracted point features can be reused across multiple viewpoints.
In contrast, the HPR method requires computing a convex hull for each viewpoint, making it less efficient.
Second, our network leverages priors learned from data, making it more robust to noise and varying point cloud densities, which enables more accurate visibility predictions in challenging scenarios.
Additionally, our method is fully differentiable and can be trained in an end-to-end manner.
This property facilitates seamless integration into other optimization or learning pipelines.

We verify the efficiency, effectiveness, and robustness of our method through extensive experiments on the large-scale ShapeNet dataset~\cite{Chang2015} and real-world scans.
Specifically, our method consistently outperforms \emph{all} previous approaches, improving visibility accuracy by 1.5\% with 81k points and by 3.7\% with 2k points on average, and by up to 6.1\% in certain categories.
Our method also accelerates computation by \emph{126 times} when the number of points is more than $200k$ compared to state-of-the-art methods.
Moreover, our network demonstrates strong generalization capabilities, performing well on general unseen shapes despite being trained on ShapeNet.
We also evaluate our method in various applications, including point cloud visualization, surface reconstruction, normal estimation, shadow casting, and viewpoint optimization.
In summary, our contributions are as follows:
\begin{itemize}[leftmargin=*,itemsep=2pt]
    \item[-] \textbf{Novel formulation}: We propose a novel learning-based approach for point cloud visibility determination, addressing limitations of previous methods and possibly inspiring new possibilities for point-based rendering.
    \item[-] \textbf{Efficient design}: Our method uses a 3D U-Net for feature extraction and a lightweight shared MLP for visibility prediction, achieving high efficiency, robustness, and generalization.
    \item[-] \textbf{Effectiveness and extensive applications}: We demonstrate the effectiveness of our method through comprehensive experiments and applications, achieving significant improvements in both visibility accuracy and computational efficiency.
\end{itemize}

\section{Related Work}\label{sec:related}

\begin{figure*}[t!]
	\centering
	\includegraphics[width=0.95\textwidth]{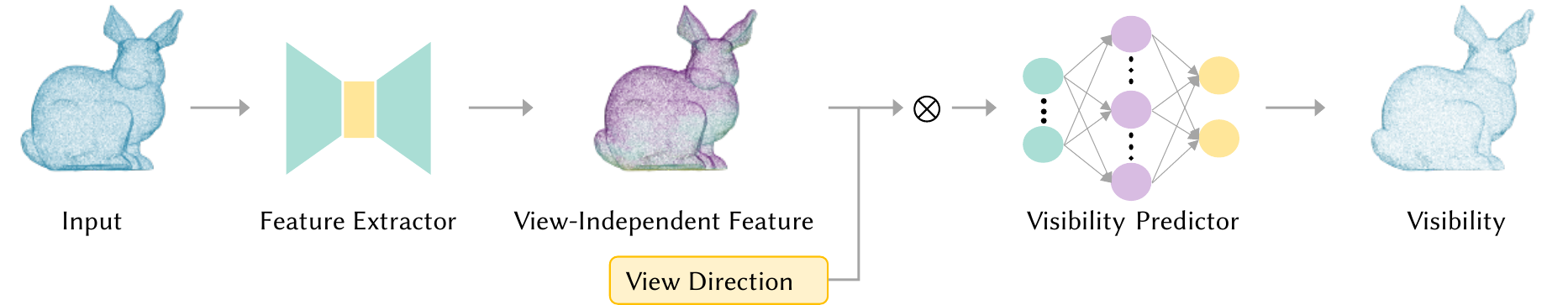} %
	\caption{Overview.
	The feature extractor is a U-Net that extracts view-independent features from the input point cloud.
	These extracted features are shared across multiple viewpoints.
	The view direction is encoded into a high-dimensional space and combined with the extracted features to produce view-dependent features. Finally, a lightweight MLP serves as the visibility predictor, estimating the visibility of each point based on the view-dependent features.
	}
	\label{fig:overview}
\end{figure*}

\paragraph{Visibility Determination}
Visibility determination is a fundamental problem in graphics and visualization~\cite{Sutherland1974,Greene1993,Bittner2003,Cohen2003,Leyvand2003}.
Traditional methods have primarily focused on determining visibility between polygons, whereas our work addresses visibility determination for point clouds, even in the absence of the surfaces they are sampled from.
This problem is particularly relevant to point-based rendering.
A common approach in point-based rendering involves expanding a point into a surfel or sphere and then splatting it to a Z-buffer~\cite{Zwicker2001,Rusinkiewicz2000,Guennebaud2004}.
However, determining the appropriate radius for the surfel or sphere is challenging, and the process often relies on point normals, which may not be available for all point clouds.
Additionally, there are also methods that use neural networks on 2D images to predict the visibility of projected points after projecting the point cloud into pixel space~\cite{Pittaluga2019,Song2021}. However, the accuracy of these methods is often limited by image resolution and point cloud density, leading to poor generalization performance on different scales of point clouds.
In contrast, our approach directly addresses the visibility problem for point clouds without expanding points into surfels or spheres and even without requiring normals.
This enables a more general and efficient solution, making it suitable for a wide range of point cloud data. \looseness=-1

\paragraph{Surface Reconstruction}
After reconstructing the surface from a point cloud, one can determine the visibility with the surface.
There exist numerous methods for surface reconstruction from point clouds,
with approaches ranging from implicit functions~\cite{Hoppe1992,Ohtake2003,Carr2001,Kazhdan2006,Kazhdan2013,Hou2022,Lin2022}, and Moving Least Squares~\cite{Alexa2003,Fleishman2003a}, to neural networks~\cite{Gropp2020,Chen2019,Park2019,Mescheder2019,Wang2023a}.
Despite significant progress, surface reconstruction remains an open challenge; refer to~\cite{Berger2017} for a comprehensive survey.
Moreover, surface reconstruction can be computationally intensive in certain scenarios.
In contrast, our method directly determines the visibility of point clouds, bypassing the surface reconstruction phase, which is not only more efficient but also well-suited for real-time applications.

\paragraph{Hidden Points Removal}
The Hidden Point Removal (HPR) method proposed by~\cite{Katz2007} are the most closely related work.
HPR directly determines the visibility of points in a point cloud by transforming the point cloud relative to a sphere centered at the viewpoint and then computing its convex hull.
A point is considered visible if it lies on the convex hull.
Subsequent works have made efforts to enhance HPR's performance on noisy and incomplete point clouds~\cite{Mehra2010}, analyze the theoretical requirements of the spherical transformation and the properties of HPR~\cite{Katz2015}, improve the efficiency of HPR by using an approximate convex hull computation~\cite{e2012efficient}, and to make HPR differentiable~\cite{katz2025hpro}.
Despite these improvements, the convex hull computation remains computationally expensive, particularly for large point clouds, while using a approximate convex hull computation can lead to degraded performance.
Additionally, HPR struggles with noise and concave regions, limiting its robustness in practical applications.
Our method leverages deep neural networks to learn priors from data and directly predict point visibility, which allows for more efficient and accurate visibility determination.

\paragraph{3D Neural Networks}
3D deep learning has made significant advancements in recent years across various applications, including 3D classification, segmentation, reconstruction, and generation.
Representative approaches include 3D voxel-based CNNs~\cite{Maturana2015,Wu2015}, sparse voxel-based CNNs~\cite{Wang2017,Graham2018,Choy2019}, and point-based neural networks~\cite{Qi2017a,Qi2017,Li2018}.
Recently, transformers have been adapted for 3D data~\cite{Guo2021,Zhao2021,Engel2021,Wu2022,Wang2023,Wu2024}.
These methods can process point clouds as input, extract point-wise features, and predict point labels.
In this paper, we build upon these advancements to tackle the visibility prediction problem of point clouds, presenting a novel application of 3D neural networks.

\section{Method}\label{sec:method}

\paragraph{Overview}
We formulate visibility determination in point clouds as a learning task and propose a deep neural network to predict point visibility.
An overview of our method is shown in \cref{fig:overview}.
Our network comprises two main components: a powerful feature extractor and a lightweight visibility predictor.
The feature extractor is a U-Net~\cite{Ronneberger2015} built upon octree-based CNNs (O-CNN)~\cite{Wang2017}, designed to extract view-independent features for each point in the input point cloud.
The visibility predictor, implemented as a multi-layer perceptron (MLP), uses these features and the view direction to predict the visibility of each point.
For each point cloud, the U-Net performs a single forward pass to extract features, which can then be reused efficiently by the visibility predictor to compute visibility across multiple viewpoints.
Unlike prior HPR methods~\cite{Katz2007,Katz2015,Mehra2010} that rely on hand-crafted priors, our approach is data-driven, learning priors directly from data.
As a result, our method is not only computationally efficient but also robust to noise, outliers, and complex concave regions, effectively addressing the limitations of HPR methods.
Next, we provide detailed explanations of the feature extractor in \cref{sec:extractor}, the visibility predictor in \cref{sec:predictor}, and training data preparation in \cref{sec:details}.

\subsection{Feature Extractor}\label{sec:extractor}

\begin{figure}[t!]
    \centering
    \includegraphics[width=0.98\linewidth]{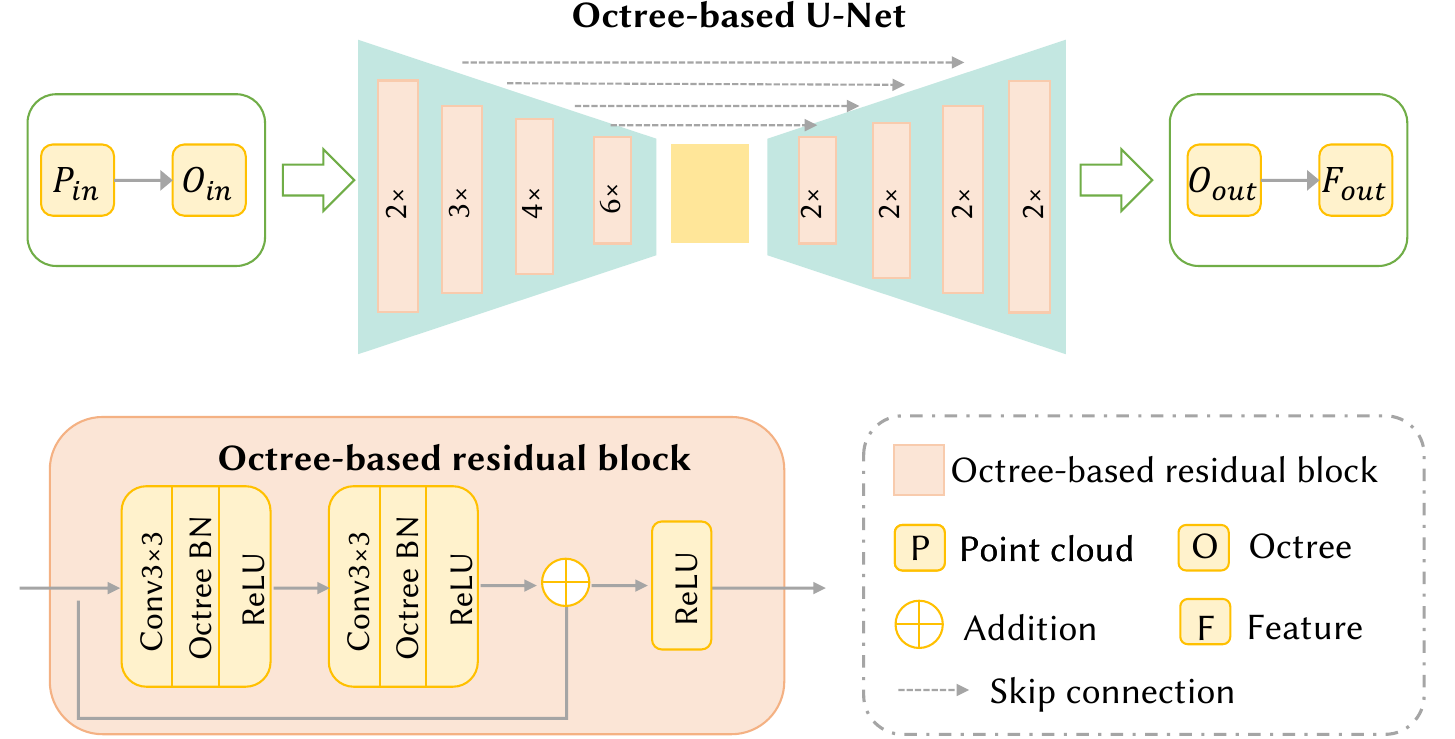} %
    \caption{Feature extractor.
    The input point cloud is converted into an octree, which serves as the input for feature extraction by a U-Net built upon O-CNN.
    The symbol $N\times$ in the figure denotes the number of residual blocks within a given stage of the network.
    The initial channel number in the residual blocks is 32, and the channel number doubles after each downsampling layer while being halved after each upsampling layer.
     }
    \label{fig:network}
\end{figure}

We adopt a U-Net architecture~\cite{Ronneberger2015} built upon O-CNN~\cite{Wang2017} as our feature extractor.
The O-CNN is an octree-based sparse 3D convolution framework, demonstrating strong performance across various 3D learning tasks.
The architecture of U-Net is illustrated in \cref{fig:network}.
It comprises a stack of residual blocks~\cite{He2016}, downsampling and upsampling layers, and skip connections.
Each residual block contains two octree-based convolutional layers, interleaved with batch normalization~\cite{Loffe2015} and ReLU activation functions.

The feature extractor takes the point cloud as input and outputs a feature vector for each point.
Given an input point cloud, we first normalize it to fit within a unit cube and convert it into an octree by recursively subdividing non-empty voxels until the maximum depth is reached.
The finest octree nodes contain the average coordinates of the points within each corresponding voxel, which serve as the input signals for the U-Net.
If normals are available, we also concatenate the average normals with the average coordinates to form the input signal.
The U-Net processes these input signals and produces a feature vector for each octree node at the finest level.
To map the features back to the original points, we employ an interpolation module defined on octrees~\cite{Yang2021a}, which splats the feature vectors to the input points.
These point-wise features are then utilized by the visibility predictor to compute visibility.
It is important to note that the feature extractor is independent of the viewpoint, allowing the extracted features to be shared across different viewpoints.


\subsection{Visibility Predictor}\label{sec:predictor}

We design a lightweight MLP as the visibility predictor to determine the visibility of each point in the point cloud.
The MLP consists of two fully connected layers, with batch normalization and ReLU activation functions.
The hidden dimensions of the MLP are set to 128 and 64, respectively.
We also experiment with a network with larger hidden dimensions, which converges faster, but the performance improvement is not significant.
The input to the MLP consists of the point-wise feature vector extracted by the feature extractor and the view direction.
Visibility determination for each point is formulated as a binary classification task.
Therefore, the output layer of the MLP has two channels, representing the scores for visible and invisible, respectively.

The view direction is represented as a unit vector on a 3D sphere.
We further encode the view direction using a set of sinusoidal functions~\cite{Tancik2020,Mildenhall2020} to help the network to learn high-frequency details for the visibility prediction.
Let the view direction be represented by a 3D vector $\mathbf{p} = (p_1, p_2, p_3)$.
Each component of the view direction is embedded as follows:
\[
    \gamma(p_i) = \left( \sin(2^0 \pi p_i), \cos(2^0 \pi p_i), \ldots,
                         \sin(2^{L-1} \pi p_i), \cos(2^{L-1} \pi p_i) \right),
\]
where $L$ denotes the number of frequencies used for encoding the view direction.
The three components of $\mathbf{p}$ are embedded separately and then concatenated.
We then multiply the embedded view direction by the extracted features from the output of the U-Net to obtain the view-dependent feature, which is subsequently fed into the MLP to predict the visibility of each point.
In the experiments, we observed that multiplying these two vectors works slightly better than concatenating them.

For the loss function, we use cross-entropy loss to train the network. The loss is computed as follows:
\[
    L = -\frac{1}{N} \sum_{i=1}^{N} y_i \log(\hat{y}_i) + (1 - y_i) \log(1 - \hat{y}_i)
\]
where $N$ is the number of points in each training batch, $y_i$ is the ground truth visibility of point $i$, and $\hat{y}_i$ is the predicted visibility.

\subsection{Data Preparation}\label{sec:details}

A remaining challenge is obtaining ground-truth visibility labels for point clouds to train the network.
Although freely available 3D datasets such as ShapeNet~\cite{Chang2015} exist, the meshes in these datasets are often of low quality, containing holes or self-intersections, and do not include visibility annotations.

To address this issue, we propose a data preparation scheme to generate synthetic training data with ground-truth visibility labels.
We begin by repairing the meshes in ShapeNet to ensure they are watertight and manifold, using the method proposed in~\cite{Wang2022}.
Next, we randomly sample $200k$ points on the surface of the mesh.
We randomly and uniformly sample viewpoints on the surface of the bounding sphere of the point cloud.
For each point in the point cloud, we connect it to the viewpoint to form a line segment and perform intersection calculations with all triangular faces. If an intersection occurs, the point is considered invisible; otherwise, it is visible.
This process is repeated for all points in the point cloud to generate the corresponding visibility labels.
Although our network is trained on synthetic data, we demonstrate in \cref{sec:result} that it generalizes well to real-world data.

\section{Results}\label{sec:result}

In this section, we validate the effectiveness and efficiency of our method through comprehensive evaluations, comparisons, and result analysis in \cref{sec:eval};
We showcase a range of applications enabled by our method in \cref{sec:app}.

\begin{table*}[t]
  \tablestyle{3pt}{1.2}
  \centering
  \caption{Quantitative comparisons on ShapeNet.
  The evaluations are conducted on point clouds with 2048, 8192, 32768 and 81920 points, denoted as $2k$, $8k$, $32k$ and $81k$, respectively. We report the mean accuracy and per-category accuracy for each method.
  The last four shaded rows indicate the results of our method when point normals are available, while the other methods cannot utilize point normals as input.}
    \begin{tabular}{lccccccccccccccccccc}
      \toprule
    Category    & Mean        & Airplane    & Bag         & Bathtub     & Bed         & Bench       & Bottle      & Cabinet     & Car         & Chair       & Display     & Lamp        & Speaker     & Pillow      & Riffle      & Sofa        & Table       & Phone   & Ship \\
    \midrule
    HPR ($2k$) & 89.3        & 87.9        & 92.7        & 88.4        & 88.0        & 86.0        & \textbf{94.7} & 91.1        & 90.2        & 87.8        & 91.1        & 88.5          & 92.5   & \textbf{94.6}        & 90.3        & 92.3        & 87.5          & 95.0        & 90.3 \\

    HPRO ($2k$) & 82.3        & 80.3        & 84.1        & 82.0        & 82.4        & 80.1        & 85.5        & 83.8        & 86.5        & 81.9        & 80.9        & 80.8        & 84.5        & 83.6        & 80.3        & 85.0        & 79.2        & 81.2        & 83.1 \\
    InvSfM ($2k$) & 39.5   & 49.8        & 43.8        & 35.2        & 35.7        & 41.3        & 38.9        & 32.3        & 28.2        & 41.0        & 44.6        & 43.5        & 32.9        & 43.8        & 48.4        & 38.2        & 41.2        & 46.4       & 40.3 \\
    \rowcolor{gray!10} Ours ($2k$)   & \textbf{93.0} & \textbf{93.6} & \textbf{94.6} & \textbf{91.9} & \textbf{90.6} & \textbf{92.1} & 94.3 & \textbf{93.4} & \textbf{92.9} & \textbf{91.2} & \textbf{94.4} & \textbf{91.8} & \textbf{92.6} & 93.3 & \textbf{95.2} & \textbf{95.2} & \textbf{92.8} & \textbf{97.0} & \textbf{93.3} \\

    HPR ($8k$) & 93.6        & 92.3        & 95.1        & 93.2        & 92.8        & 91.9        & 97.0        & 94.7        & 94.0        & 92.7        & 95.0        & 92.8        & 95.5        & 96.6        & 93.1        & 95.2        & 93.2        & 97.1        & 93.9 \\
    HPRO ($8k$) & 85.6        & 83.8        & 86.9        & 85.3        & 85.8        & 84.3        & 87.4        & 86.7        & 89.1        & 85.7        & 84.9        & 83.9        & 87.2        & 86.0        & 83.2        & 88.0        & 83.4        & 84.7       & 86.1 \\
    InvSfM ($8k$) & 52.7   & 61.0        & 57.1        & 51.1        & 50.1        & 52.6        & 54.0        & 46.0        & 44.4        & 55.3        & 55.4        & 56.8        & 46.3        & 57.6        & 58.7        & 51.8        & 52.8        & 55.6       & 54.0 \\
    \rowcolor{gray!10} Ours ($8k$)   & \textbf{96.3} & \textbf{95.9} & \textbf{96.9} & \textbf{97.0} & \textbf{95.4} & \textbf{95.0} & \textbf{97.7} & \textbf{97.8} & \textbf{96.9} & \textbf{95.2} & \textbf{97.1} & \textbf{95.0} & \textbf{97.4} & \textbf{97.4} & \textbf{96.8} & \textbf{97.7} & \textbf{96.1} & \textbf{98.6} & \textbf{96.2} \\
    HPR ($32k$) &  95.1     & 94.6      & 96.5        & 94.4        & 94.1        & 93.6        & 97.7        & 95.3        & 94.9        & 94.6        & 96.2        & 94.8        & 96.2        & 97.2        & 95.2        & 96.5        & 94.8        & 97.8        & 95.4 \\
    HPRO ($32k$) & 88.6        & 86.9        & 89.6        & 88.2        & 88.8        & 88.0        & 89.3        & 89.3        & 91.4        & 88.9        & 88.0        & 86.9        & 89.6        & 88.3        & 85.9        & 90.5        & 87.2        & 87.8       & 88.8 \\
    InvSfM ($32k$) & 65.2   & 70.3        & 69.1        & 65.3        & 64.4        & 62.2        & 69.4        & 62.8        & 62.2        & 67.1        & 64.8        & 68.2        & 64.2        & 69.0        & 67.9        & 64.9        & 63.1        & 63.5       & 66.7 \\
    \rowcolor{gray!10} Ours ($32k$) & \textbf{97.2} & \textbf{96.7} & \textbf{97.7} & \textbf{98.1} & \textbf{96.5} & \textbf{95.8} & \textbf{98.6} & \textbf{98.7} & \textbf{97.9} & \textbf{96.2} & \textbf{97.8} & \textbf{95.9} & \textbf{98.4} & \textbf{98.4} & \textbf{97.3} & \textbf{98.3} & \textbf{96.8} & \textbf{99.0}         & \textbf{97.0}         \\
    HPR ($81k$) & 96.2      & 94.7        & 96.8        & 96.9        & 95.8        & 94.8        & 98.5        & 97.1        & 96.6        & 95.7        & 97.1        & 95.6        & 97.7        & 97.9        & 95.0        & 97.2        & 96.2        & 98.3        & 96.0 \\
    HPRO ($81k$) & 87.7        & 83.7        & 87.0        & 89.4        & 89.3        & 87.9        & 86.9        & 89.5        & 92.1        & 88.3        & 86.6        & 84.6        & 89.4        & 85.9        & 81.6        & 89.1        & 87.1        & 84.9       & 87.2 \\
    InvSfM ($81k$) & 70.9   & 72.3        & 73.3        & 71.5        & 70.4        & 68.8        & 75.2        & 69.3        & 72.6        & 72.1        & 69.4        & 72.0        & 71.0        & 72.8        & 69.1        & 70.4        & 68.3        & 67.5       & 73.3 \\
    \rowcolor{gray!10} Ours ($81k$) & \textbf{97.4} & \textbf{96.8} & \textbf{97.9} & \textbf{98.2} & \textbf{96.7} & \textbf{96.1} & \textbf{98.8} & \textbf{98.8} & \textbf{98.2} & \textbf{96.4} & \textbf{97.8} & \textbf{96.1} & \textbf{98.6} & \textbf{98.6} & \textbf{97.4} & \textbf{98.4} & \textbf{97.0} & \textbf{99.0}         & \textbf{97.2}         \\
    \hline
     Ours (2k, normal) & 97.3        & 97.3        & 97.7        & 98.0          & 95.5        & 96.3        & 99.2        & 98.3        & 98.1        & 96.2        & 97.7        & 96.5        & 98.1        & 99.2        & 98.2        & 98.4        & 96.6        & 99.3        & 97.4 \\
     Ours (8k, normal) & 97.7        & 97.4        & 98.1        & 98.5        & 96.6        & 96.6        & 99.4        & 99.0          & 98.5        & 96.8        & 98.1        & 96.8        & 98.8        & 99.4        & 98.3        & 98.7        & 97.2        & 99.4        & 97.6 \\
     Ours (32k, normal) & 97.6       & 97.1        & 98.1        & 98.4        & 97.0        & 96.4        & 99.0        & 99.0        & 98.4        & 96.7        & 98.2        & 96.4        & 98.8        & 98.9        & 97.6        & 98.6        & 97.2        & 99.3        & 97.4 \\
     Ours (81k, normal) & 97.6       & 97.1        & 98.1        & 98.5        & 97.0        & 96.4        & 99.0        & 99.0        & 98.4        & 96.7        & 98.2        & 96.4        & 98.8        & 99.0        & 97.5        & 98.6        & 97.2        & 99.2        & 97.4 \\
    \bottomrule
    \end{tabular}%
  \label{tab:results}%
  \end{table*}%

\subsection{Comparisons}\label{sec:eval}

In this section, we compare our method with previous approaches to demonstrate its effectiveness, efficiency, scalability, generalization ability, and robustness.
Implementation and training details are provided in the supplementary materials.

\paragraph{Comparisons}
We compare our method with Hidden Point Removal (HPR)~\cite{Katz2007,Katz2015}, HPRO~\cite{katz2025hpro}, and InvSfM~\cite{Pittaluga2019}.
We use the visibility accuracy as the evaluation metric following previous works~\cite{katz2025hpro,Katz2015}.
We report both the mean accuracy and the per-category accuracy on the ShapeNet test set across four different point densities: 2k, 8k, 32k, and 81k points.

HPR and HPRO require parameter tuning for different point cloud densities. Their performance is sensitive to the choice of the parameter $\gamma$, which defines the kernel function for point transformation when determining point visibilities.
We tune $\gamma$ for HPR and HPRO using three groups of parameters and report the best results.
However, in real-world applications, it is impractical to fine-tune this parameter for every individual point cloud.
InvSfM employs a 2D neural network to predict point visibility in pixel space. We evaluate the pre-trained model provided by the authors using an image resolution of 512, as specified in the original paper.
In contrast, our method performs robustly across both sparse and dense point clouds without requiring any parameter tuning in the inference stage.

The quantitative results are presented in \cref{tab:results}; more detailed results for HPR and HPRO with different parameter settings are provided in the supplementary materials.
As shown in \cref{tab:results}, \emph{our approach outperforms all other methods in nearly all the cases}, particularly on sparse point clouds and complex shapes.
\begin{itemize} [leftmargin=*, itemsep=2pt]
  \item[-] For dense point clouds with 81k points, our method achieves an average accuracy of 97.4\%, outperforming the best existing method by more than 1.5\%.
  \item[-] For sparse point clouds with 2k points, our method achieves an average accuracy of 93.0\%, outperforming the best existing method by more than 3.7\%.
  \item[-] On complex categories such as \textit{table} and \textit{bench}, our method exceeds the best existing method by over 5.0\%.
\end{itemize}

\begin{figure*}[p]
    \centering
    \includegraphics[width=\textwidth]{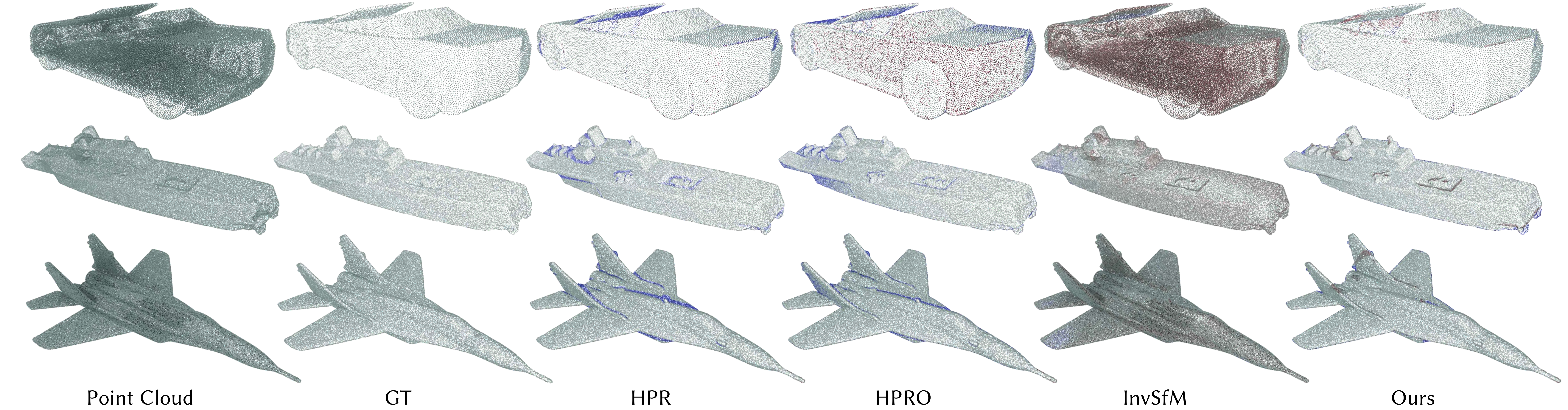} %
    \caption{Results comparisons.
    From left to right: input point cloud, ground-truth visibility, results from HPR, HPRO, InvSfM, and our method. Each point cloud contains 100k points. Red points indicate false positives, while blue points indicate false negatives.
    Compared to other methods, our approach produces more accurate visibility predictions, particularly in regions with high curvature.
    }
    \label{fig:results}
\end{figure*}

These results highlight the significant performance improvements of our method, demonstrating its effectiveness and robustness compared to other approaches.
Visual comparisons on the ShapeNet dataset are shown in \cref{fig:results}.
It can be observed that HPR and HPRO often produce false negatives in regions with high curvature, while InvSfM frequently generates false positives in occluded regions.
Our method effectively captures fine geometric details and provides more reliable visibility predictions in challenging regions.


\paragraph{Visibility with Normals}
Point normals are available in some real-world point clouds.
Our method can be extended to incorporate point normals as additional input features.
We report the visibility accuracy of our method under this setting.
As shown in the last four rows of \cref{tab:results}, incorporating normals further boosts our method’s performance, surpassing other methods by a considerable margin.
However, the other approaches cannot leverage point normals as input, highlighting a distinct advantage of our method in utilizing additional geometric information for improved accuracy.

\begin{figure}[htbp]
  \centering
  \includegraphics[width=\columnwidth]{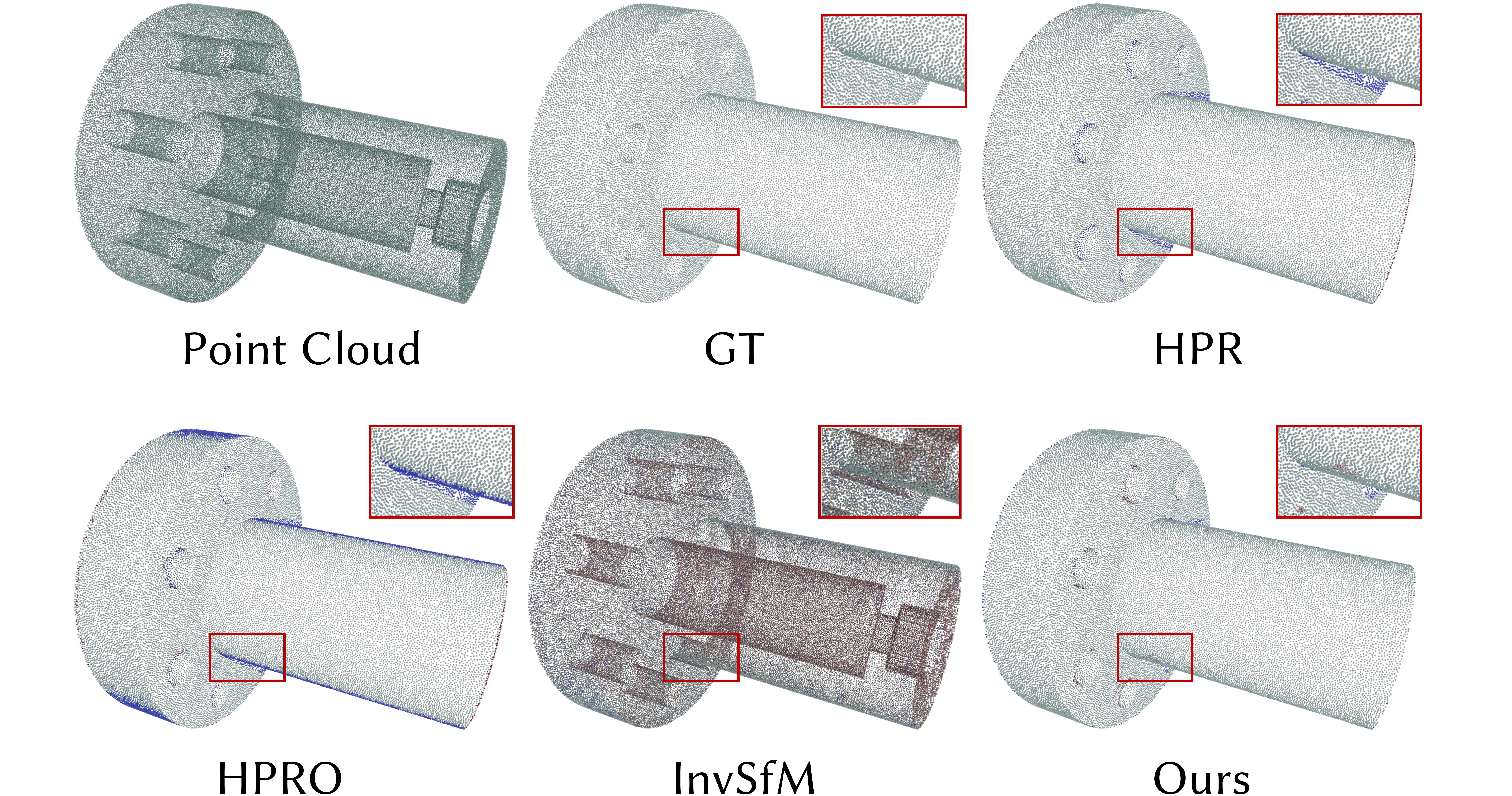} %
  \caption{Results on ABC Dataset.}
  \label{fig:abc_fig}
\end{figure}

\paragraph{Generalization}
We design experiments to evaluate the generalization ability of our method on unseen shapes.
Specifically, we train the network using only 13 categories from the ShapeNet dataset, as opposed to the 18 categories used in previous experiments, and test it on the remaining 5 categories.
All other training settings are kept unchanged.
The results are summarized in \cref{tab:generalization}.
We observe that our method maintains consistent high accuracy on unseen shapes, with only a slight decrease of less than 0.3\% compared to training on all 18 categories.
This demonstrates the strong generalization capability of our approach in handling unseen shapes.
We also report cross-dataset results on the ABC dataset~\cite{Koch2019}, using the model trained on the ShapeNet dataset, as shown in \cref{tab:abc_generalization}.
The accuracy on the ABC dataset is slightly lower than that on the ShapeNet dataset, likely due to differences in the point cloud distributions between the two datasets.
Some point clouds in the ABC dataset contain multiple objects, which our method cannot handle well.
Even so, our method achieves accuracies of 89.0\% and 92.5\% on point clouds with 2k and 81k points, respectively, still outperforming other methods.
The results are shown in \cref{fig:abc_fig}.

\begin{table}[t]
  \tablestyle{10pt}{1.2}
  \caption{Generalization on unseen shapes.
  \emph{Seen} means these categories are included in the training set, while \emph{Unseen} means these categories were not used during training.}
  \begin{tabular}{lccccc}
  \toprule
          & Bag         & Bed         & Bottle      & Pillow      & Phone \\
    \midrule
    Unseen ($2k$)      & 94.6        & 90.4        & 94.1        & 93.1        & 96.9 \\
    Seen ($2k$)        & 94.6        & 90.7        & 94.3        & 93.2        & 97.0 \\
    Unseen ($8k$)      & 96.8        & 95.3        & 97.5        & 97.3        & 98.5 \\
    Seen ($8k$)        & 96.9        & 95.4        & 97.7        & 97.4        & 98.6 \\
    Unseen ($32k$)     & 97.6        & 96.5        & 98.5        & 98.4        & 98.9 \\
    Seen ($32k$)       & 97.7        & 96.5        & 98.6        & 98.4        & 99.0 \\
    Unseen ($81k$)     & 97.8        & 96.7        & 98.7        & 98.7        & 98.9 \\
    Seen ($81k$)       & 97.9        & 96.7        & 98.8        & 98.7        & 99.0 \\

  \bottomrule
  \end{tabular}%
  \label{tab:generalization}%
\end{table}%

\begin{table}[t]
  \tablestyle{17pt}{1.2}
  \caption{Accuracy comparisons on the ABC dataset.
  Our method is trained on the ShapeNet dataset and directly evaluated on the ABC dataset.
  }
  \begin{tabular}{lccc}
    \toprule
    Point Number     & 2k     & 8k      & 32k \\
    \midrule
    HPR      & 88.0  & 90.5   & 92.2  \\
    HPRO     & 80.4  & 83.1 & 85.6 \\
    InvSfM & 40.6  & 46.6   & 56.1 \\
    Ours     & \textbf{89.0}  & \textbf{91.7}   & \textbf{92.4}\\
    \bottomrule
    \end{tabular}%
  \label{tab:abc_generalization}%
  \end{table}%


\begin{figure*}[p]
    \centering
    \includegraphics[width=\textwidth]{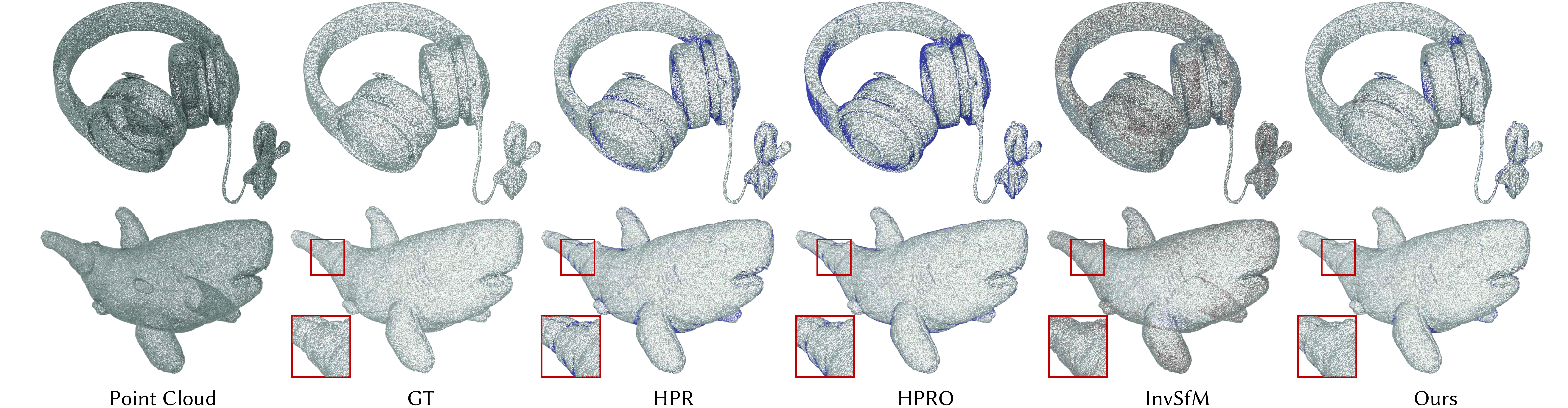} %
    \caption{Results on real-world point clouds.
     From left to right: input point cloud, ground-truth visibility, results from HPR, HPRO, InvSfM, and our method. Red points indicate false positives, while blue points indicate false negatives.
    }
    \label{fig:real_data}
\end{figure*}

\begin{figure*}
    \centering
    \includegraphics[width=\textwidth]{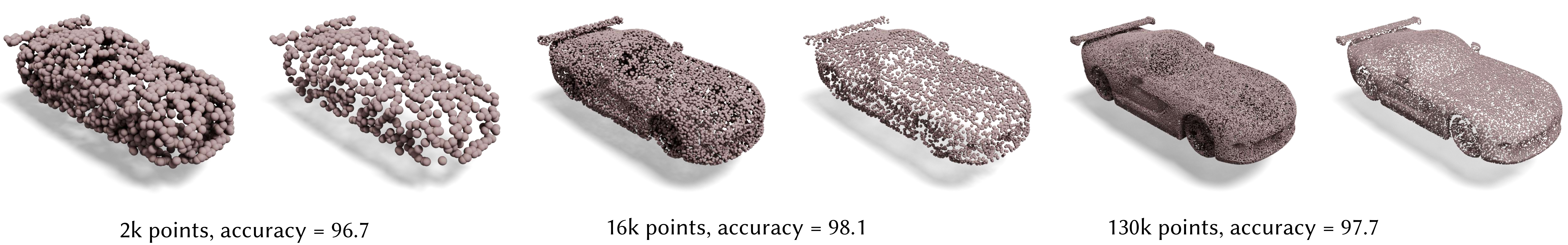} %
    \caption{Scalability. We test our method on point clouds with different scales ranging from 2048 to 130k points, without retraining. Our method demonstrates excellent scalability, with no significant drop in accuracy as the scale increases.}
    \label{fig:scalability_fig}
  \end{figure*}

\begin{figure*}
    \centering
    \includegraphics[width=\textwidth]{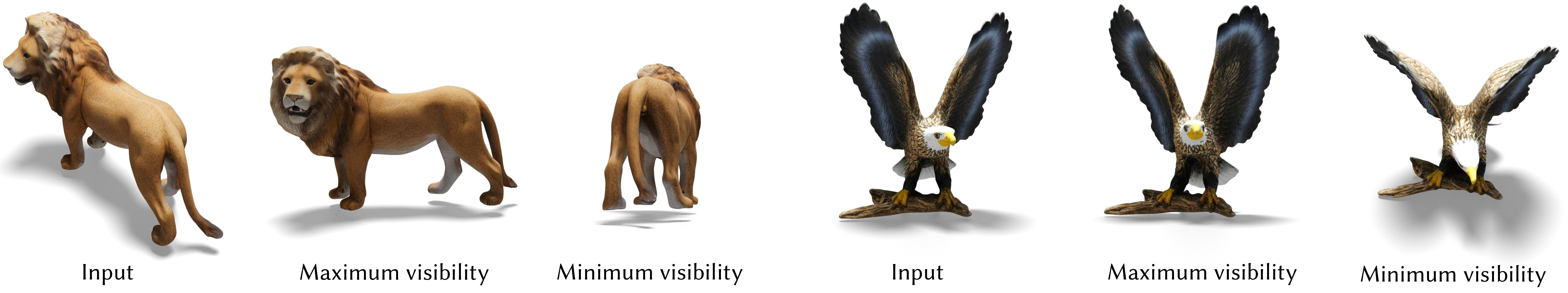} 
    \caption{View optimization.
    We take the vertices of the mesh as input to extract features with our network, and optimize for 300 iterations to find the maximum and minimum view.
    }
    \label{fig:best_view}
\end{figure*}

\paragraph{Scalability}
The octree effectively resamples input point clouds, producing approximately consistent features across different sampling densities, thereby ensuring scalability.
We train a model on point clouds ranging from 2k to 8k points, and evaluate its performance on point clouds with $16k$ and $130k$ points.
The results are shown in \cref{fig:scalability_fig}.
The accuracy of our method remains nearly constant as the scale of the point cloud increases, demonstrating excellent scalability.
This property is particularly advantageous for real-world applications, where point clouds often vary in size and density.
However, HPR and HPRO require manual parameter tuning to accommodate different point densities, and InvSfM is also sensitive to the number of points in the point cloud.


\begin{table}[t]
  \tablestyle{9.5pt}{1.2}
  \caption{Robustness.
  The evaluations are conducted on point clouds with 2048 ($2k$) and 8192 ($8k$) points, with noise levels of 1\% and 2\%.
  Our method demonstrates strong robustness, consistently outperforming HPR.
  }
    \begin{tabular}{lcclcc}
    \toprule
    Noise Level      & 1\%    & 2\%    &Noise Level      & 1\%    & 2\%\\
    \midrule
    HPR ($2k$)     & 81.4        & 79.2     & HPR ($32k$)     & 84.7        & 81.3\\
    HPRO ($2k$)     & 80.8        & 78.9     & HPRO ($32k$)     & 82.6        & 79.1\\
    InvSfM ($2k$) & 39.2        & 39.1     & InvSfM ($32k$) & 64.1        & 62.7\\
   \rowcolor{gray!10} Ours ($2k$)    & \textbf{88.9}        & \textbf{87.8}     & Ours ($32k$)    & \textbf{93.3}        & \textbf{91.4}\\
    HPR ($8k$)     & 83.6        & 80.8     & HPR ($81k$)     & 83.5        & 79.3 \\
    HPRO ($8k$)     & 82.3        & 79.5     & HPRO ($81k$)     & 79.5        & 76.0\\
    InvSfM ($8k$) & 52.1        & 51.6     & InvSfM ($81k$) & 69.7        & 68.0\\
   \rowcolor{gray!10} Ours ($8k$)    & \textbf{91.8}        & \textbf{90.4}     & Ours ($81k$)    & \textbf{94.0}        & \textbf{91.6}\\
    \bottomrule
    \end{tabular}%
  \label{tab:noisy}%
\end{table}%

\begin{figure}[htbp]
    \centering
    \includegraphics[width=\columnwidth]{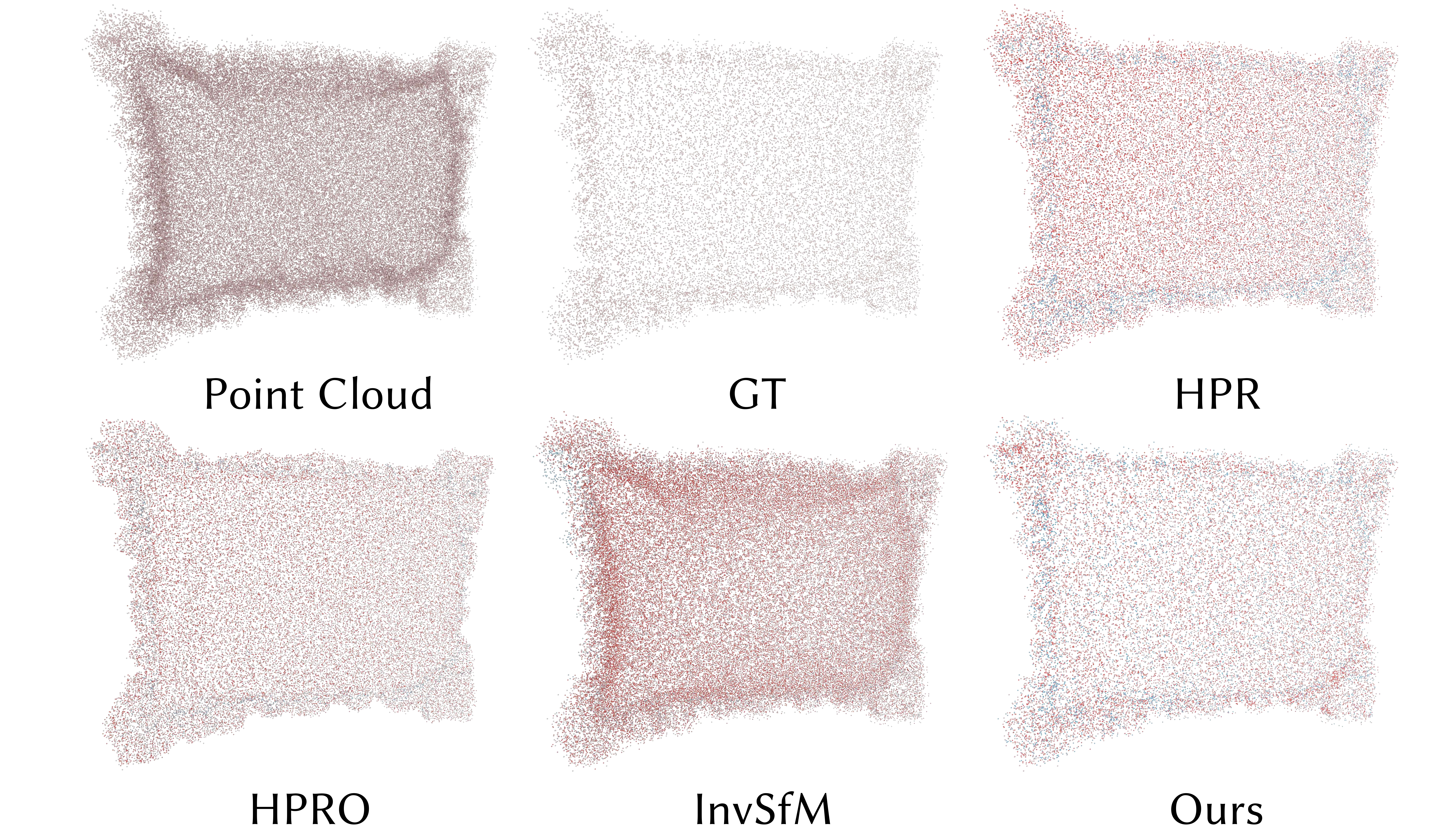} %
    \caption{Results on noisy point clouds.
    The noise levels is 0.01.
    The red and blue points represent the false positives and false negatives, respectively.
    Our method significantly improves the accuracy, demonstrating its robustness.}
    \label{fig:noise_results}
\end{figure}

\paragraph{Robustness}
We evaluate the robustness of our method on noisy and incomplete point clouds.
For noisy point clouds, we add uniformly distributed noise of varying magnitudes to each point, following the approach in~\cite{Mehra2010}.
Specifically, we train our network on point clouds with 2\% noise and evaluate its performance on point clouds with noise levels of 1\% and 2\%.
The noise level refers to the upper bound of the noise magnitude, expressed as a percentage of the bounding box size.
The results are presented in \cref{tab:noisy}.
Our method demonstrates strong robustness to noise, outperforming other approaches by more than 7\% in accuracy across both noise levels.
Visual results are shown in \cref{fig:noise_results}.
Compared to other methods, our approach effectively reduces both false positives and false negatives, particularly at higher noise levels.

\begin{figure}[htbp]
    \centering
    \includegraphics[width=0.93\columnwidth]{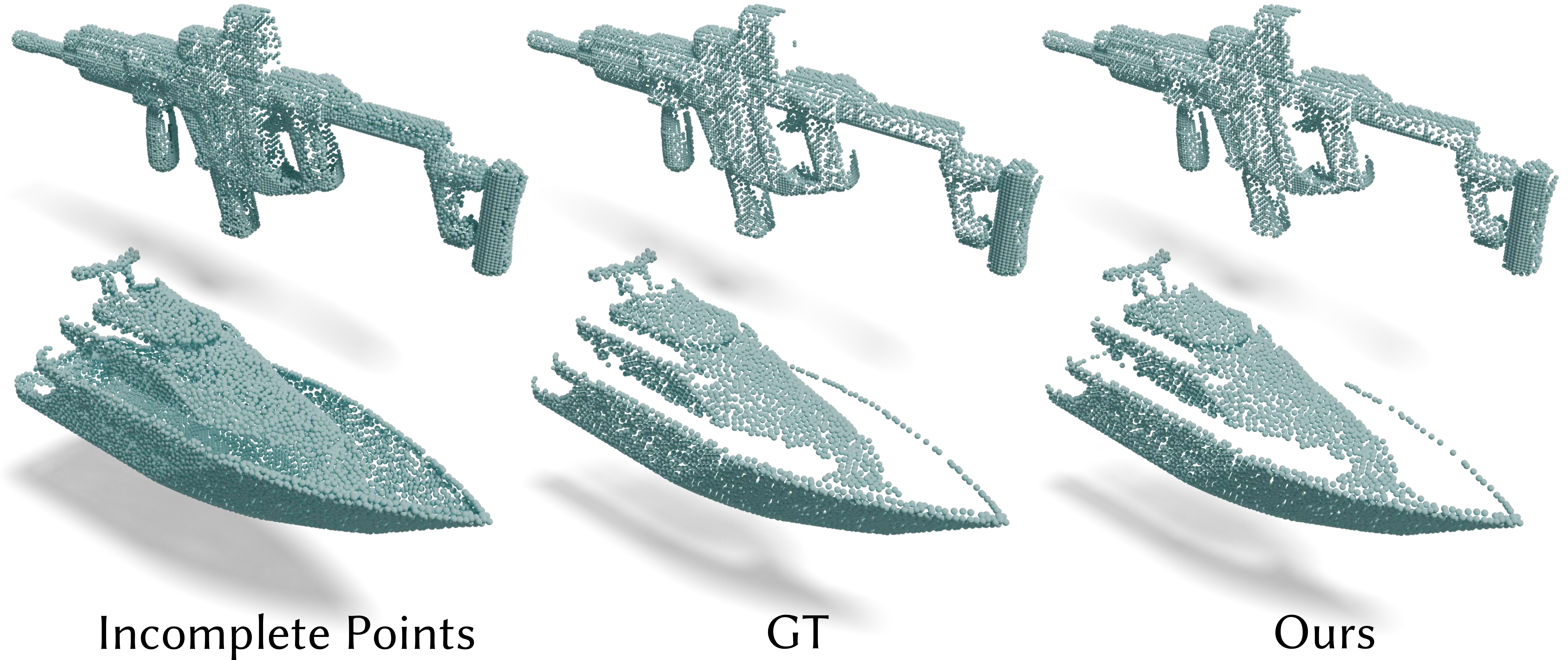} %
    \caption{Results on incomplete point clouds.
    Our model is trained on ShapeNet, it can accurately predict the visibility for incomplete point clouds.
    }
\label{fig:incomplete_results}
\end{figure}

For incomplete point clouds, we randomly select three viewpoints and remove points that are not visible from any of them.
We directly apply the model trained on complete point clouds to these incomplete inputs.
As shown in \cref{fig:incomplete_results}, even with incomplete point clouds, our method accurately predicts point visibility.
This indicates that our method effectively learns shape priors from the training data, enabling it to generalize well to incomplete data.

\begin{figure}[t]
  \centering
  \includegraphics[width=\linewidth]{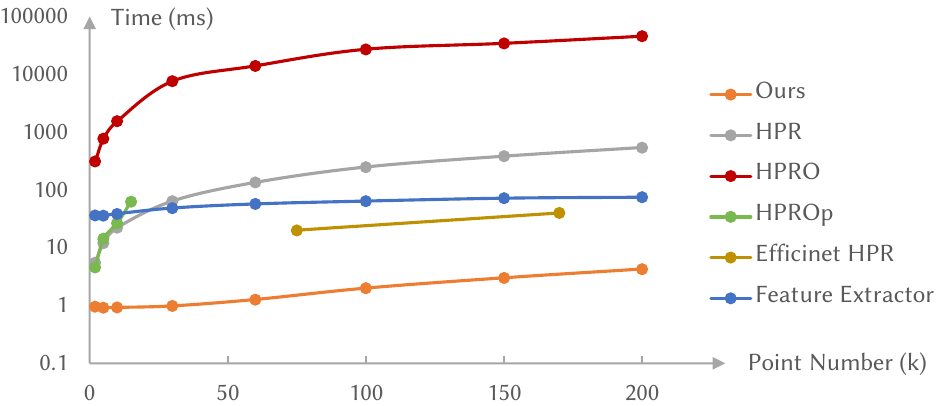} %
  \caption{Efficiency comparison.
  HPROp is a parallel implementation of HPRO with significantly higher memory usage. With 24GB of memory, we can only test it on point clouds containing up to 15k points.
  The results of Efficient HPR is from~\cite{e2012efficient}. 
  Feature Extractor is the time consumption of our U-Net, which is forwarded once for each point cloud.
  }
  \label{fig:efficiency}
\end{figure}

\paragraph{Efficiency}
One of the key advantages of our method is its efficiency.
The HPR method requires constructing a convex hull for each viewpoint, with an average time complexity of $\mathcal{O}(n \log n)$ and a worst-case complexity of $\mathcal{O}(n^2)$, where $n$ is the number of points in the point cloud.
The Efficient HPR method~\cite{e2012efficient} improves upon this by leveraging a GPU-based implementation to approximate the convex hull, reducing the average time complexity to $\mathcal{O}(n \log n / m)$, where $m$ depends on the number of GPU cores.
In contrast, our method computes visibility through a simple feed-forward pass of a lightweight two-layer MLP after extracting features from the point cloud with the octree-based U-Net.
As a result, the time complexity of our method is effectively $\mathcal{O}(n / m)$.

The comparison of time consumption for point clouds of varying sizes is presented in \cref{fig:efficiency}.
As the number of points increases, the inference time of our visibility predictor (the MLP) grows approximately linearly beyond 50k points.
Meanwhile, the time required for the feature extractor (the U-Net) converges to a constant, as the number of nodes in the octree is bounded by its maximum depth.The preprocessing time of our feature extractor (the U-Net) is 75ms for 200k points on single L40 GPU with memory cost less than 5GB, which is forwarded once for each point cloud, still much faster than HPR.
When the point cloud exceeds 200k points, our method significantly outperforms the HPR method, achieving a speedup of up to \emph{126 times}.
Specifically, our method processes a point cloud of this size in just 4.27\,ms, compared to 541.5\,ms for HPR.

\paragraph{Real-world Point Clouds}
We evaluate our method on real-world point clouds from the Google Scanned Objects dataset~\cite{Downs2022}.
The results are shown in \cref{fig:real_data}.
These real-world point clouds exhibit significantly greater geometric complexity than those in the ShapeNet dataset.
Despite being trained solely on ShapeNet, our method generates accurate visibility predictions on these real-world examples, demonstrating strong robustness and generalization capabilities.
Compared to HPR, our approach consistently achieves better performance, particularly in high-curvature regions, highlighting its effectiveness in handling complex and diverse shapes.


\begin{table}[t]
  \tablestyle{16pt}{1.2}
  \caption{Effects of the octree depth.
  The first row presents the time consumption of the feature extractor in milliseconds, the second row shows the corresponding accuracy.
  }
  \begin{tabular}{lcccc}
    \toprule
    Depth     & 6     & 7      & 8       & 9   \\
    \midrule
    Time      & 30.7  & 35.6   & 49.3    & 75.0 \\
    Accuracy  & 95.7  & 97.1   & 97.4    & 97.3 \\
    \bottomrule
    \end{tabular}%
  \label{tab:Depth}%
\end{table}%

\subsection{Ablation Study and Discussions}
In this section, we conduct ablation studies to analyze the impact of key design choices in our model.

\paragraph{Octree Depth}
We evaluate the effect of octree depth on the performance and efficiency on the ShapeNet dataset.
The results are summarized in \cref{tab:Depth}.
A small octree depth leads to a coarse representation of the point cloud, which may result in a loss of important geometric details, while a large octree depth increases the computational cost of extracting features. In our experiments, we set the octree depth to 8, which provides a good balance between accuracy and efficiency.

\paragraph{Transformer-based Feature Extractor}
Transformers are generally considered more effective at capturing global relationships.
To evaluate their performance in our task, we replaced the OCNN-based feature extractor with OctFormer~\cite{Wang2023}.
The results are presented in \cref{tab:transformer}.
The accuracy difference between the two architectures does not exceed 1\%.
This suggests that the CNN-based U-Net has sufficient capacity to capture global context for the visibility prediction task.

\begin{table}[t]
  \tablestyle{14pt}{1.2}
  \caption{{Comparisons with transformer-based feature extractor. OctFormer performs similarly to OCNN-based U-Net.}}
  \begin{tabular}{lcccc}
    \toprule
    Point Number     & 2k     & 8k      & 32k       & 81k   \\
    \midrule
    OCNN     & 93.0  & 96.3   & \textbf{97.2}    & \textbf{97.4} \\
    Octformer  & \textbf{93.8}  & \textbf{96.4}   & 97.0    & 97.1 \\
    \bottomrule
    \end{tabular}%
  \label{tab:transformer}%
  \end{table}%

\begin{table}[t]
  \tablestyle{12pt}{1.2}
  \caption{Ablation study on position embedding and U-Net architecture.}
  \begin{tabular}{lcccc}
    \toprule
    Point Number     & 2k     & 8k      & 32k       & 81k   \\
    \midrule
    w/o position embedding & 92.3  & 95.7   & 96.7    & 96.9 \\
    5-layer Unet & \textbf{93.0}  & \textbf{96.3}   & \textbf{97.2}    & \textbf{97.4} \\
    Ours     & \textbf{93.0}  & \textbf{96.3}   & \textbf{97.2}    & \textbf{97.4} \\
    \bottomrule
    \end{tabular}%
  \label{tab:ablation}%
  \end{table}%
\looseness=-1


\paragraph{Position Embedding and Unet Architecture}
We evaluate the effect of position embedding and the U-Net architecture on the performance of our method. As shown in \cref{tab:ablation}, using position embedding improves the performance of our method by 0.5\% on average, from 96.9\% to 97.4\% on the ShapeNet dataset. As for the U-Net architecture, We tested a deeper U-Net (5 layers), which increased the parameter count from 40 M to 130 M, yet the test accuracy remained the same, which is identical to the current result.

\paragraph{Additional Metrics}
We computed the false-negative/positive rates on ShapeNet, measuring the proportions of visible points predicted as invisible, and vice versa. Our method achieved 4.3\% and 2.0\%, respectively, compared to HPR's 9.2\% and 1.1\%. Notably, our false-negative rate is 4.9\% lower than HPR's.



\begin{figure}[t]
  \centering
  \includegraphics[width=\linewidth]{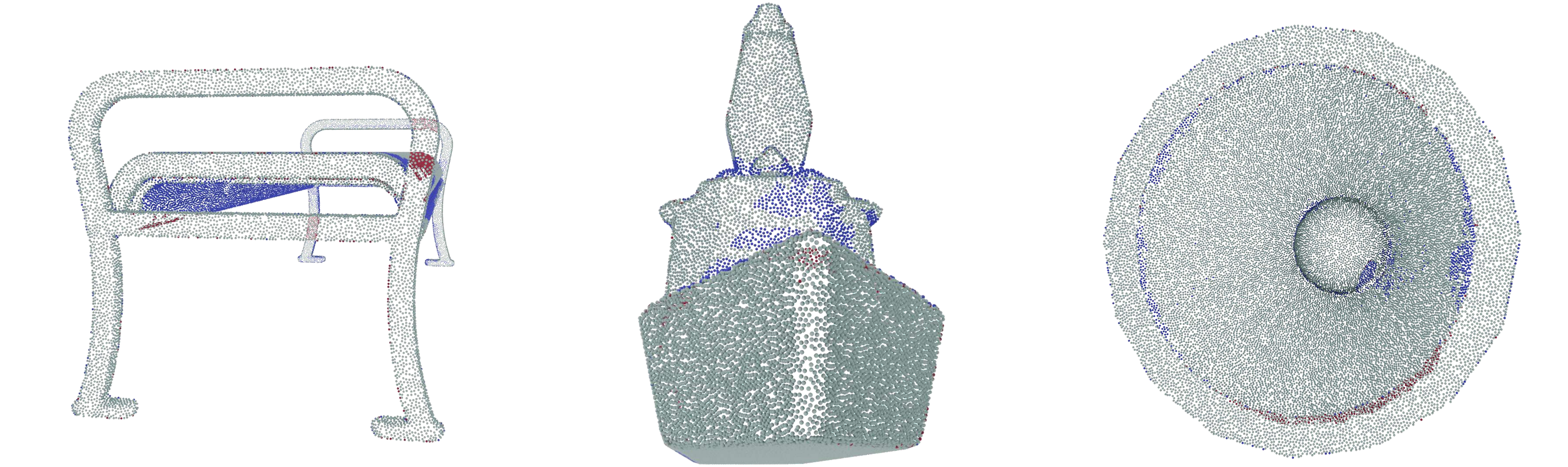} %
  \caption{Failure cases on ShapeNet. }
  \label{fig:failcase}
\end{figure}

\paragraph{Failure cases}
Our method sometimes encounters failure cases on thin structures, large curvature concave surfaces, and multi-layer occlusions. Some examples are shown in \cref{fig:failcase}. This may require improvement on the network or the training strategy.

\looseness=-1

\begin{figure}[t]
  \centering
  \includegraphics[width=\linewidth]{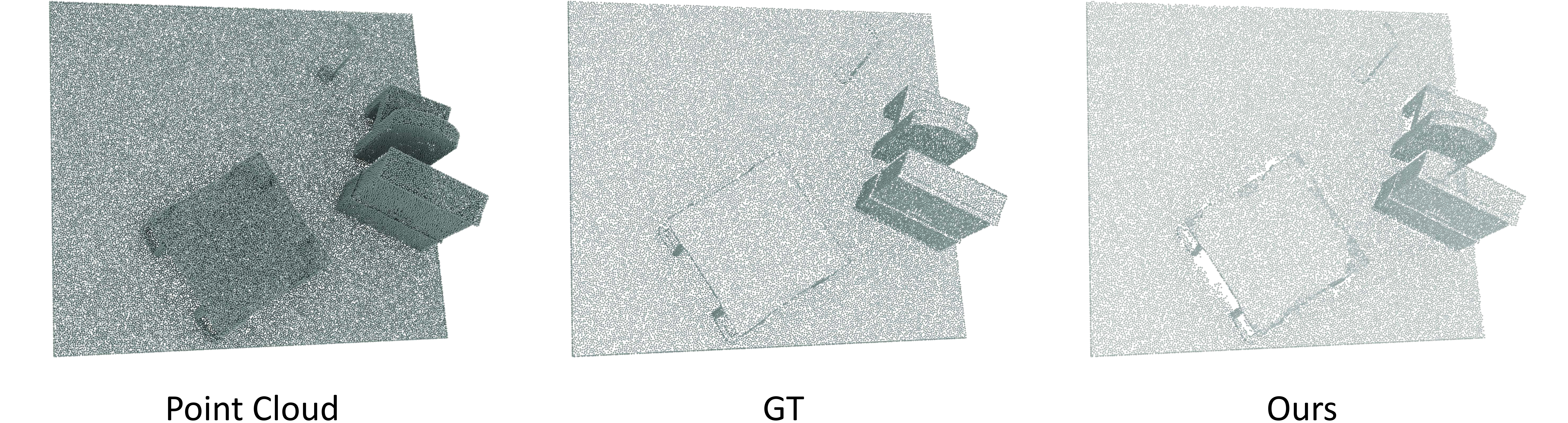} %
  \caption{Results on scene-level point clouds. Left: input point clouds. Right: visibile points predicted by our method.}
  \label{fig:scene}
\end{figure}

\paragraph{Scene-level Results}
We evaluate our model on a scene dataset composed of multiple ShapeNet objects\cite{Peng2020}, with 500 training and 125 testing scenes. Using the ShapeNet-trained model, we achieve 94.3\% accuracy. Fine-tuning on the scene dataset further improves accuracy to 95.1\%. Results are shown in \cref{fig:scene}. Our method performs slightly worse in predicting occlusions between objects, which still requires further improvement.

\looseness=-1

\subsection{Applications}\label{sec:app}

In this section, we showcase a range of applications enabled by our method, including point cloud visualization, best view selection, view-dependent surface reconstruction, normal estimation, and shadow casting.

\begin{figure}[H]
    \centering
    \includegraphics[width=\columnwidth]{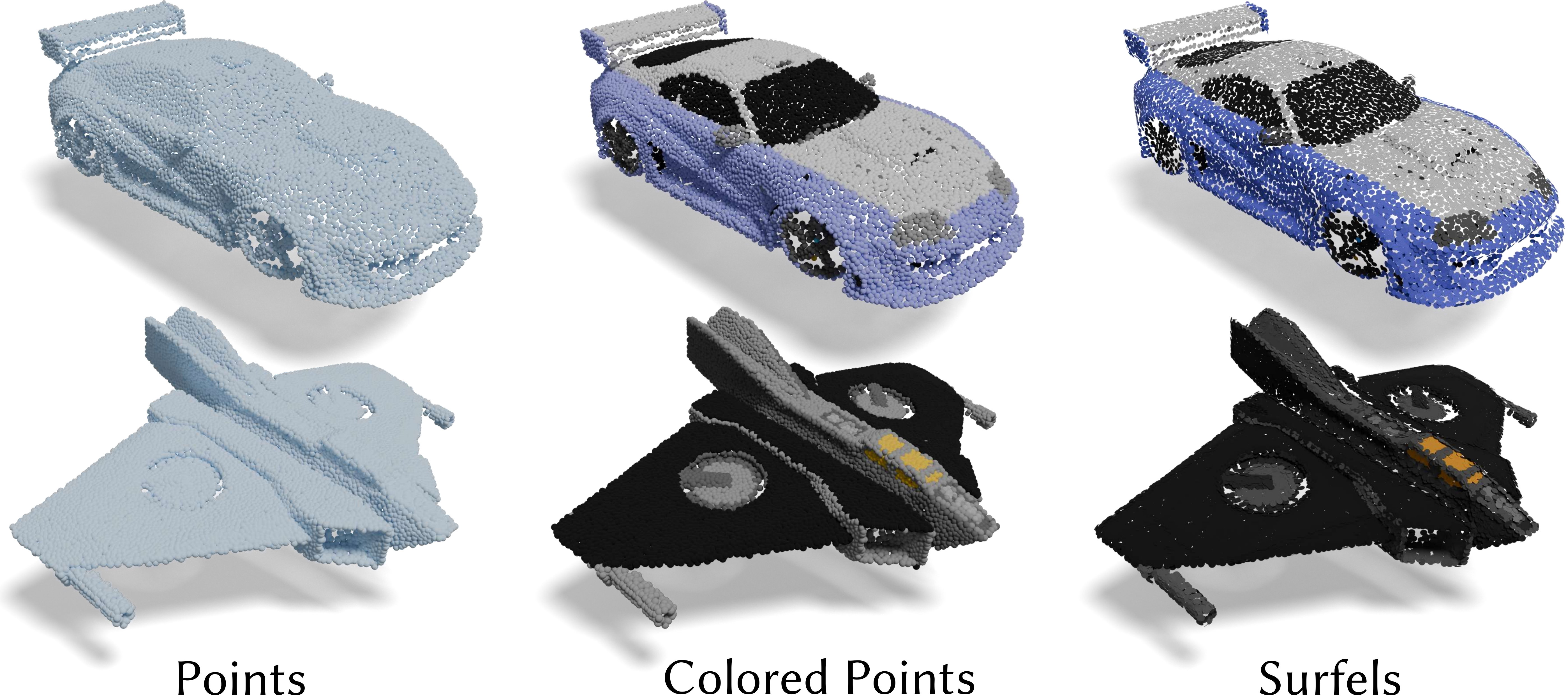} 
    \caption{Point cloud visualization. The left column shows the results of our method, while the middle and right columns are rendered with color and surfels, respectively.}
  \label{fig:visualize}
  \end{figure}

\paragraph{Point Cloud Visualization}
A direct application of our method is point cloud visualization.
By rendering only the visible points, our approach effectively removes occluded regions, resulting in a clear and informative representation of the shape.
This is particularly advantageous for point clouds containing color or normal information, as it significantly enhances the clarity and interpretability of the rendering results.
Several examples are shown in \cref{fig:visualize}.

\begin{figure}[htbp]
  \centering
  \includegraphics[width=0.93\columnwidth]{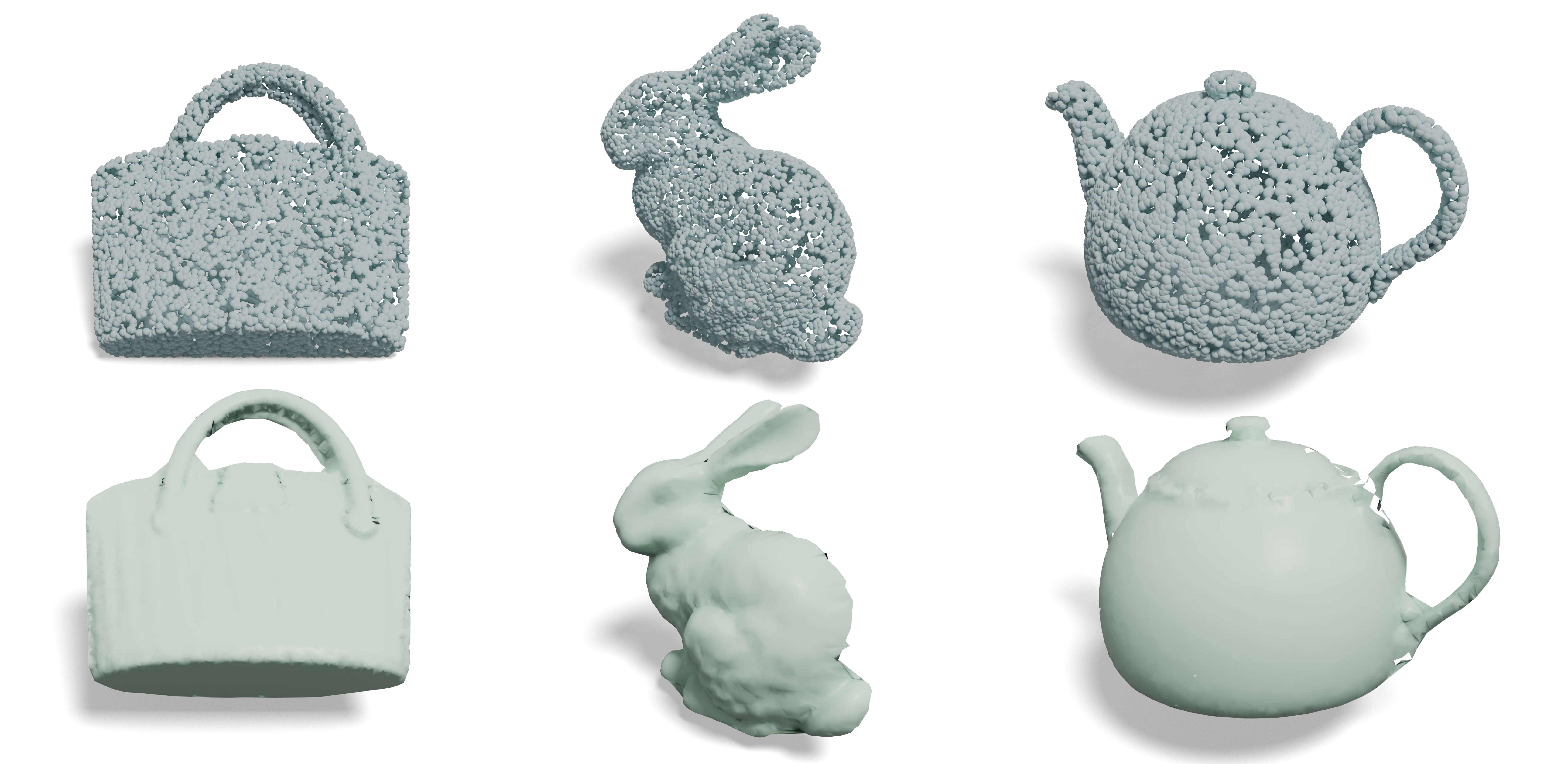} 
  \caption{View-dependent Surface Reconstruction.}
  \label{fig:reconstruction}
\end{figure}

\paragraph{View-dependent Surface Reconstruction}
Our method enables surface reconstruction of a 3D shape from a point cloud, even in the absence of point normals.
After determining the visibility of each point, we project the visible points into the view space and apply Delaunay triangulation to establish the connectivity among points.
The resulting triangulation is then mapped back to 3D to generate a view-dependent mesh.
To maintain the visual coherence of the reconstructed surface, we discard triangles whose edge lengths exceed a predefined threshold.
The results are presented in \cref{fig:reconstruction}.
\looseness=-1

\begin{figure}[htbp]
  \centering
  \includegraphics[width=\columnwidth]{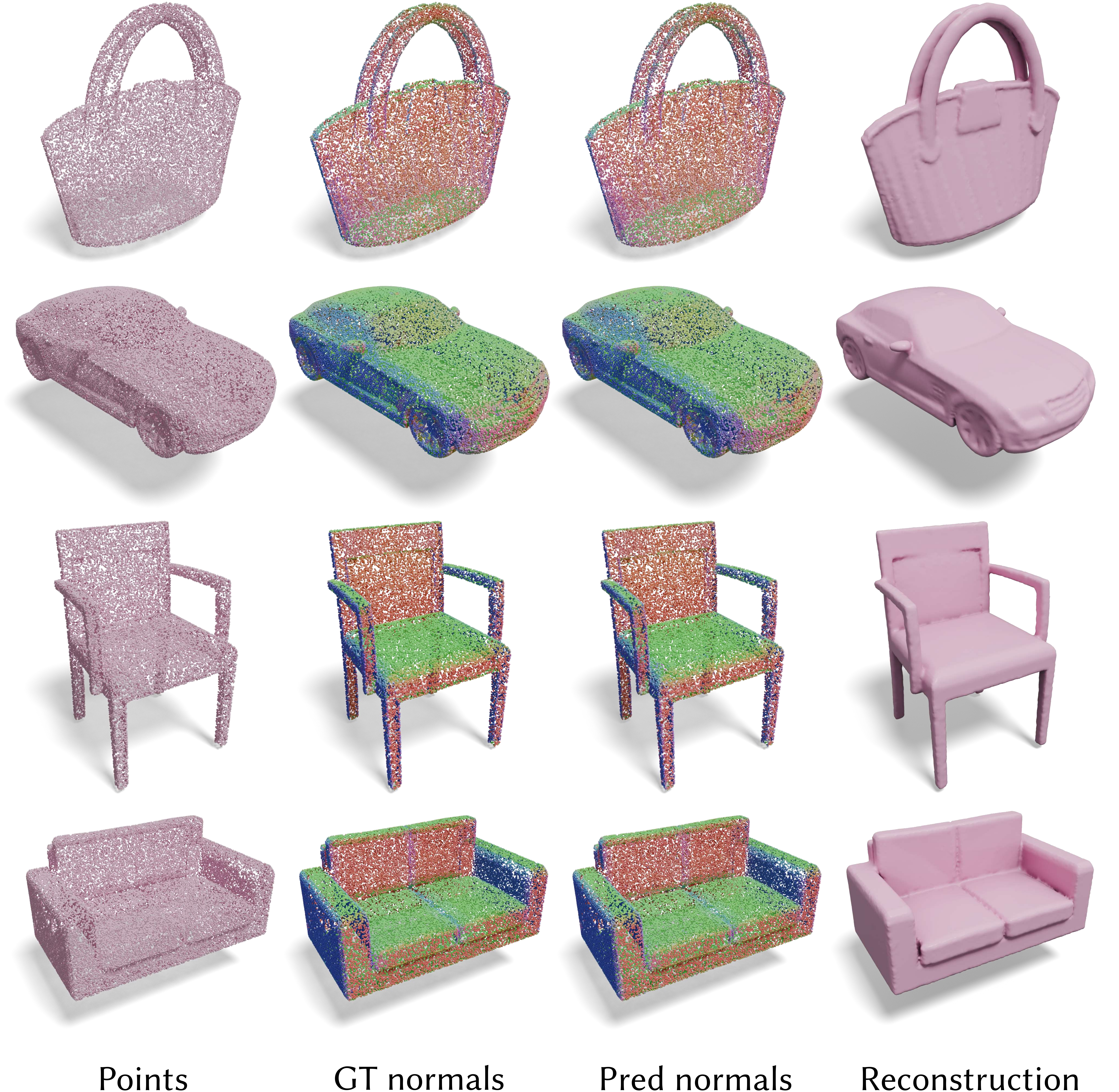} 
  \caption{Normal estimation and reconstruction. We predict and aggregate the normals from 26 fixed viewpoints, then reconstruct the surface by Poisson surface reconstruction. The cosine similarity between the predicted normals and the ground truth normals is all above 0.93.}
  \label{fig:normal}
\end{figure}

\paragraph{Normal Estimation}
Following view-dependent surface reconstruction, we can estimate the normal of each point by averaging the normals of the triangles that contain it.
We uniformly sample 26 viewpoints around the point cloud and compute the average of the normal estimations across these views.
For points that are not visible from any of the sampled views, we propagate the normal from the nearest visible point.
The comparison results are shown in~\cref{tab:normal-est-quan}.
Using the resulting predicted normals, we perform surface reconstruction via Poisson surface reconstruction~\cite{Kazhdan2006}.
The results are presented in~\cref{fig:normal} and the quantitative results are presented in~\cref{tab:surface-recon-quan}.

\begin{table}[t]
  \tablestyle{10pt}{1.2}
  \caption{Quantitative pointcloud normal prediction result comparison between PCPNet~\cite{Paul2017} and our method. We tested using three sets of data: PCPNet's test data (from its provided dataset) \textbf{A}, our test data (from the ShapeNet dataset) \textbf{B}, and test data unseen during training (from the COSEG dataset) \textbf{C}.}
  \begin{tabular}{lcccc}
    \toprule
    Dataset & Method & Mean Radian Error & Mean Angle Error \\
    \midrule
    A       & PCPNet & 0.156            & 6.63             \\
    A       & Ours   & 0.156            & 9.14             \\
    B       & PCPNet & 0.126            & 7.23             \\
    B       & Ours   & 0.062            & 3.55             \\
    C       & PCPNet & 0.085            & 4.84             \\
    C       & Ours   & 0.059            & 3.39             \\
    \bottomrule
    \end{tabular}%
  \label{tab:normal-est-quan}%
  \end{table}
\looseness=-1

\begin{table}[t]
  \tablestyle{10pt}{1.2}
  \caption{Quantitative mesh surface reconstruction accuracy results conducted on a non-training dataset(COSEG).}
  \begin{tabular}{lcccc}
    \toprule
    Category & Chamfer Distance & Hausdorff Distance & F1-score \\
    \midrule
    Chairs   & $5\times10^{-5}$          & 0.0228            & 1.000  \\
    Fourleg  & $7\times10^{-5}$          & 0.0241            & 1.000  \\
    Guitars  & $5\times10^{-5}$          & 0.0182            & 0.999  \\
    Lamps    & $5\times10^{-5}$          & 0.0166            & 1.000  \\
    \bottomrule
    \end{tabular}%
  \label{tab:surface-recon-quan}%
  \end{table}
\looseness=-1



\begin{figure}[htbp]
  \centering
  \includegraphics[width=\columnwidth]{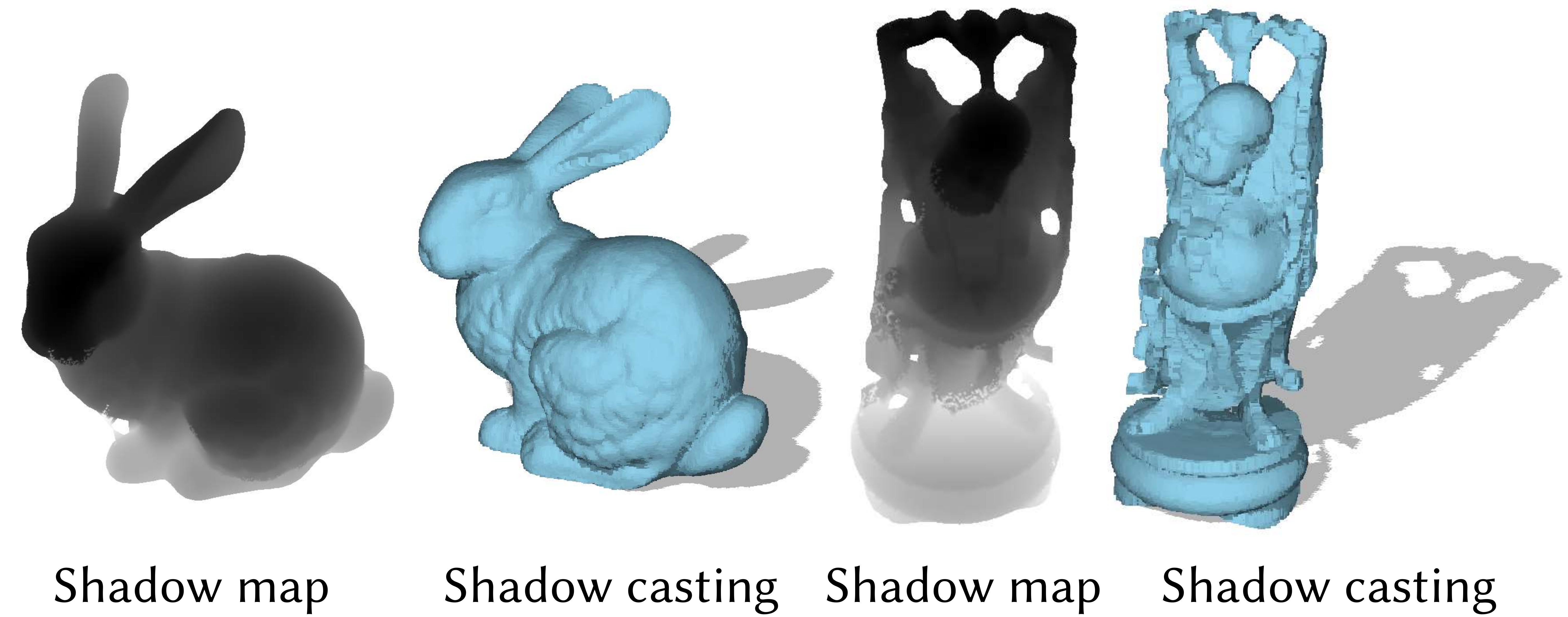} 
  \caption{Shadow casting. The first and third images show the depth map under the light source, which are used as the shadow maps. The second and fourth images show the rendering results with shadows.}
  \label{fig:Shadow}
\end{figure}

\paragraph{Shadow Casting}
Shadow casting is essential for enhancing the realism of 3D scenes in computer graphics~\cite{Hasenfratz2023}.
Using our method, shadows can be seamlessly incorporated into point clouds.
To render shadows, we treat the light source as the viewpoint and determine shadowed and illuminated regions using a z-buffer in object space.
Our approach predicts the visible points directly from the perspective of the light source.
We then construct the connectivity of these visible points using Delaunay triangulation.
By interpolating their depth values, we generate a depth map from the light source's perspective, which serves as a shadow map for rendering.
This process enables accurate identification and rendering of shadowed regions.
The results are shown in \cref{fig:Shadow}.

\begin{figure}[H]
  \centering
  \includegraphics[width=\linewidth]{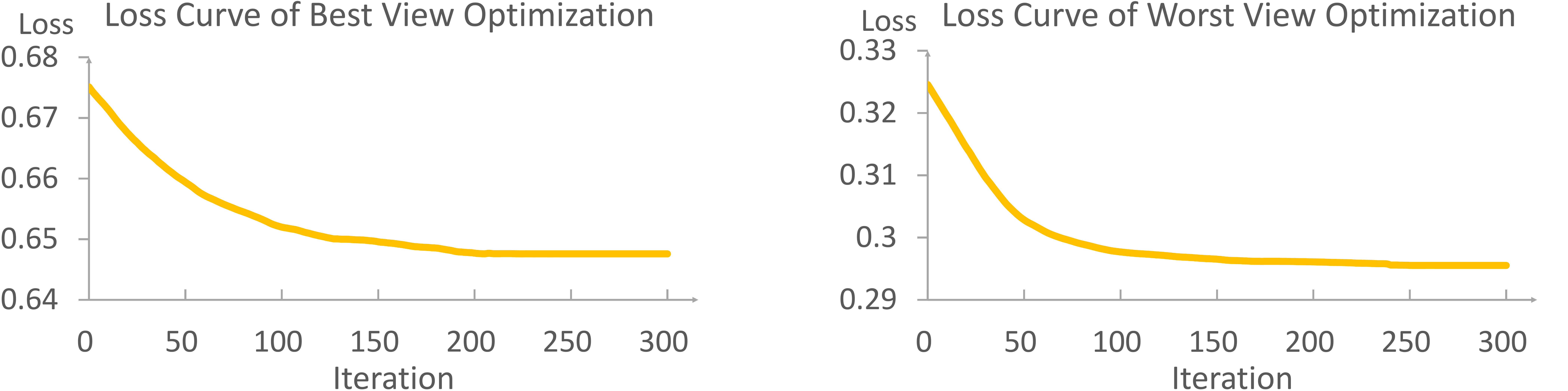} %
  \caption{Loss curve of best and worst view optimization.}
  \label{fig:best_view_curve}
\end{figure}

\paragraph{View Selection}
View selection is an important problem in computer graphics and visualization~\cite{Leifman2016}.
As our approach is differentiable, we can optimize the viewpoint directly by minimizing the invisibility score of the point cloud defined as: \looseness=-1
\[
    \mathcal{L}(v) = \frac{1}{N} \sum_{i=1}^{N} o_i(v),
\]
where $v$ is the viewpoint, $o_i(v)$ is the invisibility score predicted by the network for the $i$-th point, and $N$ is the total number of points in the point cloud.
By minimizing this loss while keeping the network parameters fixed, we can effectively identify a viewpoint that maximizes visible details and minimizes occlusions, resulting in a more comprehensive view of the shape.
Conversely, by maximizing the invisibility score, we can determine the viewpoint that yields the least visible details.
The results are shown in \cref{fig:best_view}, and the loss curve during optimization is presented in \cref{fig:best_view_curve}.

\section{Conclusions} \label{sec:conclusion}

This paper introduces a novel approach for visibility determination in point clouds by framing the problem as a learning task.
Utilizing an octree-based U-Net for feature extraction and a shared multi-layer perceptron (MLP) for visibility prediction, our method offers significant improvements over traditional techniques in both accuracy and computational efficiency.
Extensive experiments on ShapeNet and real-world datasets demonstrate the robustness and generalization capabilities of our method.
Moreover, our method supports a variety of applications, including surface reconstruction, normal estimation, shadow casting, and viewpoint optimization.
We hope that our work will inspire future research in visibility determination and facilitate the development of more efficient and robust methods for point-based rendering and point cloud processing.

Despite these advancements, there are several limitations and future works for improvements.
First, our model focuses on single objects with viewpoints located outside the visual hull.
Extending it to handle scene-level visibility prediction or multi-object scenarios would be a valuable direction for future research.
Second, our method relies on ground-truth visibility labels for training, which may not always be available in real-world scenarios.
Future work could explore semi-supervised or self-supervised learning strategies to reduce dependency on labeled data.
Third, to better capture partial occlusions or transparency effects, extending our approach to model probabilistic visibility rather than binary visibility is a promising avenue.
Finally, incorporating additional data augmentation techniques and large-scale training strategies could further enhance generalization, especially in complex real-world settings.

\begin{acks}
  This work was supported in part by National Natural Science Foundation of China (Grant No. 62371409) and Beijing Natural Science Foundation (Grant No. 4244081).
  We also thank the anonymous reviewers for their invaluable feedback.
\end{acks}

\clearpage
\bibliographystyle{ACM-Reference-Format}
\bibliography{src/reference}



\end{document}